\documentclass{ws-procs961x669}            

\newcommand{\bra}[1]{\langle{#1}|}
\newcommand{\ket}[1]{|{#1}\rangle{}}

\begin{document}
\bibliographystyle{ws-procs961x669}

\title{Progress and open problems in evolutionary dynamics}

\author{Richard A.~Neher}
\address{Biozentrum, University of Basel, Switzerland}

\author{Aleksandra M.~Walczak}
\address{CNRS and LPTENS, PSL, Paris, France}

\begin{abstract}
Evolution has fascinated quantitative and physical scientists for decades: 
how can the random process of mutation, recombination, and duplication of genetic information generate the diversity of life? What determines the rate of evolution? Are there quantitative laws that govern and constrain evolution? Is evolution repeatable or predictable?
Historically, the study of evolution involved classifying  and comparing species, typically based on morphology. 
In addition to phenotypes on the organismal and molecular scales, we now use whole genome sequencing to uncover not only the differences between species, but also to characterize genetic diversity within-species in unprecedented detail.
This diversity can be compared to predictions of quantitative models of evolutionary dynamics.
Here, we review key theoretical models of population genetics and evolution along with examples of data from lab evolution experiments,  longitudinal sampling of viral populations, microbial communities and the studies of immune repertoires. 
In all these systems, evolution is shaped by often variable biological and physical environments.
While these variable environments can be modelled implicitly in cases such as host-pathogen co-evolution, the dynamic environment and emerging ecology often cannot be ignored. 
Integrating dynamics on different scales, both in terms of observation and theoretical models, remains a major challenge towards a better understanding of evolution.
\end{abstract}


Over the centuries, scientists have been cataloging the diversity of life, and through systematic classification based largely on morphological features, uncovered its hierarchical organization. 
This hierarchical structure prompted the idea that organisms evolve by passing on their phenotypic traits to their offspring with possible modification. 
On much shorter time scales, plant and animal breeding demonstrated that phenotypic traits can be dramatically altered by consistently selecting for desired types. The advent of genetics and molecular biology in the 20th century established the molecular basis for the inheritance and modification of phenotypic traits.
In the 20th century mathematical frameworks were developed to describe genetic diversity within species (population genetics and quantitative genetics) and between species (phylogenetics).
However, until recently molecular data to test specific predictions and assumptions have been lacking.  The advent of relatively cheap high-throughput sequencing has changed this situation profoundly. 
The elegant theoretical models, however, often fail to describe the now abundant data sets in a quantitative manner.

These data sets are either observations of natural populations or are obtained from laboratory evolution with microbes or other rapidly reproducing organisms, e.g.~flies. 
Laboratory evolution experiments with microbes have the great advantage over observational studies in animals or plants that population turnover and the spread of mutations happens on observable time scales.
Resulting time dependent datasets are much more powerful at  differentiating between models than snapshots of natural populations. 
The only natural `measurably evolving` populations where observations beyond snapshots can be made used to be pathogenic bacteria, RNA viruses (influenza virus and HIV) and certain cancer types. 
But time resolved data of commensal and environmental bacterial populations are starting to emerge \citep{garud_evolutionary_2017,bendall_genome-wide_2016,zhao_adaptive_2017}.
Strong selection and rapid mutation are also at play during somatic hypermutation of B-cell receptors.
Genetic diversity of entire microbial populations or immune receptors can be characterized by high-throughput sequencing and allows  genome wide tracking of mutations at high resolution.

These experiments and observational studies have shown that microbes adapt rapidly to new environments, even if these environments are kept as simple and stable as possible. There are ample opportunities for adaptation and populations are typically diverse with many similar trait modifications competing against each other. 
Such rapid and diverse responses are at odds with classical population genetics theory, which assumes that adaptive changes are rare and occur sequentially with periods of stasis between so-called selective sweeps.

Furthermore, models typically assume an externally specified environment in which every member of the population competes with everybody else for a common resource. However, even in the simplest experiments that sought to approximate this idealized model, the populations rapidly split into different types that feed on secondary metabolites or specialize in growth at different nutrient densities. 
In other words, even simple environments seem to rapidly develop internal ecology. Ecological theory, on the other hand, typically ignores evolution and assumes that species are static monomorphic entities that interact with each other. Putting evolution into the context of ecology and {\it vice versa} is the major scientific frontier for the field as it goes forward.

Here, we review theoretical models of population genetics and discuss recent data that suggest that the assumptions of many established theoretical models are not met. 
We discuss theoretical developments that aim at addressing some of the conflicts, and present an overview of the wide open questions that arise when evolution and ecology meet. 

\section*{Traditional population genetics}
Population geneticists and quantitative geneticists in the 20th century attempted to develop these qualitative ideas of heritability, mutation, and selection into a quantitative description of evolution. 
The models studied in the 20th century can be broadly classified into (i) deterministic dynamical systems, (ii) phenomenological quantitative genetics models, and (iii) stochastic models of frequencies genetic variants segregating within species. 
Quantitative genetics models describe the response of diverse populations to selection for particular phenotypes without reference to the genetic determinants of the traits \citep{falconer_introduction_2003}.
Patterns of genetic diversity and distributions of variant frequencies, on the other hand, require probabilistic models of the processes that introduce and remove genetic variation.

The two prominent models for genetic diversity are the backwards in time ``coalescent'' \citep{Kingman:1982p28911} or a forward in time diffusion approach \cite{Kimura:1955p36295}. 
The Kimura diffusion equation describes the distribution $P(x,t)$ of the frequency $x$ of a single genetic variant $A$ subject to genetic drift and selection in a (haploid) population of size $N$.
If individuals with and without variant $A$ have on average $1+s$ and $1$ offspring, respectively, and the variance in offspring number is $\sigma^2$, the distribution of the variant frequency evolves according to
\begin{equation}
    \frac{\partial P(x,t)}{dt}  = \left[-s\frac{\partial}{\partial x} x(1-x)  + \frac{\sigma^2}{2N}\frac{\partial^2}{\partial x^2}x(1-x)\right]P(x,t).
\end{equation}
The first term on the right accounts for selection with strength $s$; the second term accounts for demographic stochasticity known as \emph{genetic drift}. 
Genetic drift results from the variance $\sigma^2$ in offspring number. 
In a population of size $N$, the frequency $x$ of a variant therefore has a diffusion constant $\frac{\sigma^2}{N}x(1-x)$.
This relation between offspring number variance and frequency diffusion assumes that fluctuations are uncorrelated across individuals and generations (non-heritable). 
In most populations, however, many processes affect variant frequencies in an undirected manner but their effects are correlated in time and space.  
If these correlations are weak and decay sufficiently rapidly, they can be accounted for by introducing an effective population size $N_e$. 
The latter is often orders of magnitudes smaller than estimates of the census population size, suggesting that genetic drift in a narrow sense is typically irrelevant.

When fluctuations are correlated over many generations, their multiplicative effect can be dramatic and qualitatively change the stochastic dynamics. In this case, different models outside the universality class of the Kimura diffusion equation are needed.
One such example are the effects of other loci in the genome that are under directional selection -- known as \emph{hitch-hiking} or \emph{genetic draft}: by consistently changing frequencies in the same direction, a selective sweep at a nearby locus can result in macroscopic perturbations to neutral diversity \citep{gillespie_genetic_2000}.

\begin{figure}[tb]
    \centering
    \includegraphics[width=0.48\columnwidth]{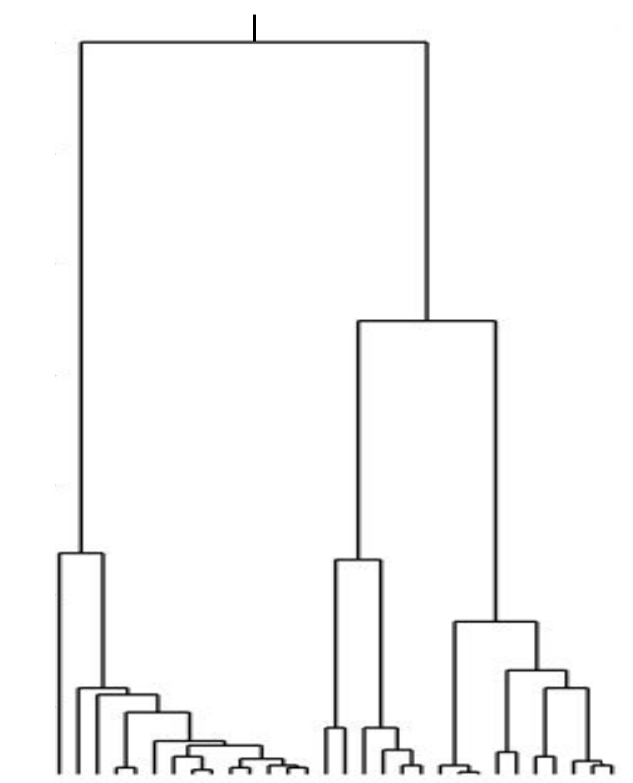}
    \includegraphics[width=0.48\columnwidth]{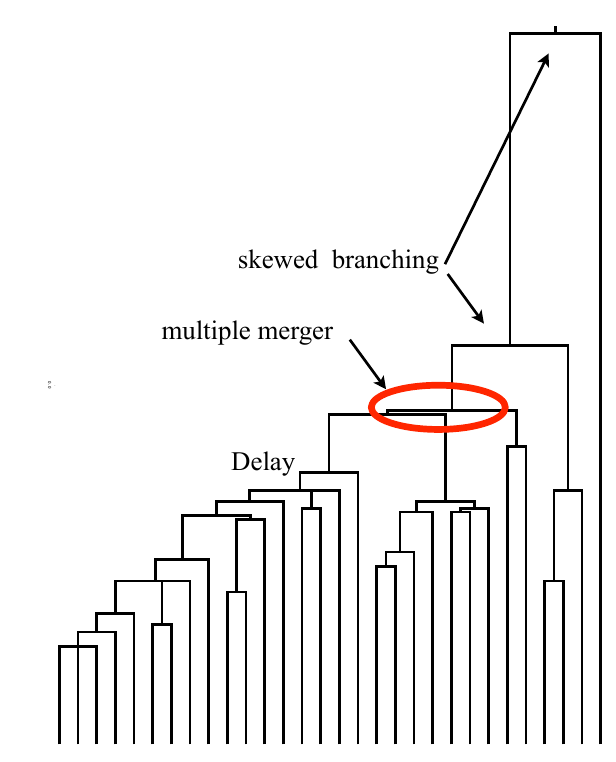}
    \caption{{\bf Coalescent trees in neutral and adapting populations.} The left panel shows a typical Kingman coalescent tree. Most mergers happens close to the present (bottom) and the remaining lineages tend to split the populations into similarly sized families. The tree on the right was sampled from a simulation of an adapting population. At first, little coalescence is happening until ancestral lineages have moved to the high fitness tail. Here, competition between exponentially growing clones results in approximate multiple mergers and skewed branching.}
    \label{fig:coalescent_trees}
\end{figure}

Another popular framework to analyze neutral diversity ($s=0$ for all variants) is the Kingman coalescent, which models the genealogy of the population backward in time: any pair of lineages merges at random with rate $k(k-1)\sigma^2/2N$, where $k$ is the number of remaining lineages. 
A typical tree generated by the Kingman process in shown in Fig.~\ref{fig:coalescent_trees}.
The time to the most recent common ancestor ($T_{MRCA}$) of a large population is on average $2N/\sigma^2$, where $\sigma^2$ is again the variance in offspring number. 
The coalescent can be used to predict diversity at many genomic loci and is easy to simulate under a range of demographic models. 
Diffusion and coalescent descriptions are two dual ways of looking at the same processes, but some questions are more conveniently formulated in one framework than the other.

The basic coalescent model assumes that all individuals are equivalent, i.e., all variation is neutral and there is no population structure. 
These assumptions can be relaxed by introducing a structured coalescent where individuals have types that restrict who can merge with whom \citep{hudson_gene_1990}. 
While Kimura's diffusion equation readily accommodates positive selection at a single locus, it is challenging to incorporate positive selection into the Kingman coalescent \citep{krone1997ancestral}.

A different framework that has been explored in some depth is that of a monomorphic population evolving in a static landscape that assigns fitness to all possible genotypes.
This approach assumes that mutations are rare enough and fitness differences between neighboring genotypes large enough that the population is monomorphic most of the time. 
This dynamics results in a Boltzmann distribution with the inverse population size and fitness playing the role of energy and temperature, respectively \citep{iwasa_free_1988,Berg2004,sella_application_2005}.

\section*{Confronting theories with data}
While sequencing a single bacterial genome was a major undertaking 20 years ago, a small lab can now sequence hundreds of bacterial genomes in a few days. 
Thousands of animal, plant, or parasite genomes have been sequenced and observed patterns of genetic diversity in these populations have been compared to predictions of patterns of genetic diversity \citep{corbett2015natural}.
Using these data, scientists have estimated demographic parameters such as the $T_{MRCA}$, historical population size changes and migrations \citep{li2011inference,schiffels2014inferring,weissman2017minimal,gutenkunst2009inferring}. 
The observed patterns are qualitatively consistent with predictions in the sense that large populations tend to harbor more genetic diversity and that there are many more rare than common variants.

However, essentially all sequence data from eukaryotes are static snapshots and our ability to learn dynamical properties from such data is limited. 
In the best case, inference from static data yields estimates of model parameters relative to an intrinsic time scale of the system, e.g.~the above mentioned coalescent time scale (or effective population size). 
More often, static data are unable to differentiate between models and seemingly good fits of simple models hide the actual dynamics.
Any inference from static data is necessarily an average over history. 
Different ways of calculating expectation values of the same parameters average over different past intervals and can give contrasting results.
For example, Bergland et al.\citep{bergland2014genomic} surveyed genetic variation in Drosophila populations over several annual cycles and found oscillations of large amplitudes suggesting strong selection pressures (10\% or more) that vary with seasons. 
By contrast, inference from static data suggests selection coefficients of 1\% or less \citep{sella2009pervasive}.
This discrepancy is just one example of how inference from static data can lead to misleading results because the dynamics are not identifiable from static data alone.

To study evolutionary processes in real-time rather than inferring them from snapshots, high throughput sequencing technologies have been applied, for example, to the long term evolution experiment (LTEE) with E.~coli conducted by Richard Lenski and colleagues \citep{barrick_genome_2009}, the tracking of millions of lineages over short times intervals \citep{levy_quantitative_2015}, and sequencing of serially sampled HIV populations \citep{zanini_population_2016}, or the global surveillance of influenza viruses \citep{gisrs_who_2017}. 

Evolution experiments have the advantage over observational studies that the environments can be controlled and replicated, but the environments used to propagate the populations are typically simple, artificial and stable. As a result, the majority of adaptive mutations that are observed are loss of function mutations, that is inactivation of proteins and pathways that are not necessary in the lab environment.
This mode of adaptation is not necessarily representative of evolution in the wild.

Outside the lab, rapid evolution can be directly observed in populations of pathogenic RNA viruses such as HIV or influenza. 
These viruses continuously evade a co-evolving immune system.
The Global Influenza Surveillance and Response System (GISRS) \citep{gisrs_who_2017} collects thousands of influenza virus samples. 
The common ancestor of all circulating A/H3N2 viruses is typically only 3 years in the past and in any given year two randomly sampled HA sequences differ on average at $\sim 10$ positions. 
This rapid turnover and diversification is driven by increasing human immunity against circulating viruses. New variants with different antigenic properties emerge and reinfect previously immune individuals  \citep{petrova_evolution_2017}.

Similar dynamics can be observed in HIV populations in infected humans. 
Shortly after infection, the adaptive immune systems targets the virus and escape mutations quickly spread \citep{ganusov_fitness_2011}. 
Typically, a few strongly selected mutations evading cytotoxic T-lymphocyte responses spread during the first 3 month of infection, followed by a multitude of more weakly selected escape and reversion mutations. 
While segments of the influenza virus genome are propagated asexually, HIV recombines by crossing over (template switching of the reverse transcriptase \citep{levy2004dynamics}). 
Recombination is rapid enough that different regions of the genome show profoundly different dynamics\citep{neher_recombination_2010}. 
Surface proteins exhibit rapid turnover similar to global influenza virus dynamics, while enzymes often slowly accumulate diversity over many years \citep{zanini_population_2016,shankarappa_consistent_1999}.

Good et al \citep{good_dynamics_2017} recently sequenced 120 samples of each of the 12 lines of the LTEE that evolved in a constant environment for about 60,000 generations.
They discovered that multiple lineages co-existed in these cultures for many years.
Instead of there being a simple environment with a single fitness optimum, the bacteria partitioned the system into niches creating an environment in which multiple types can coexist.

These three examples, together with many other recent observations of microbial evolution, have demonstrated that diverse populations, in which many variants compete vigorously are the rule rather than the exception. 
Furthermore, the environments in which microbes evolve change, either because they coevolve with a host or because they generate their own ecology.
A quantitative description of these populations requires theoretical frameworks beyond single locus diffusion theory, neutral coalescent models, or static fitness landscapes.

\section*{Traveling waves models of rapid adaptation}
\begin{figure}[tb]
    \centering
    \includegraphics{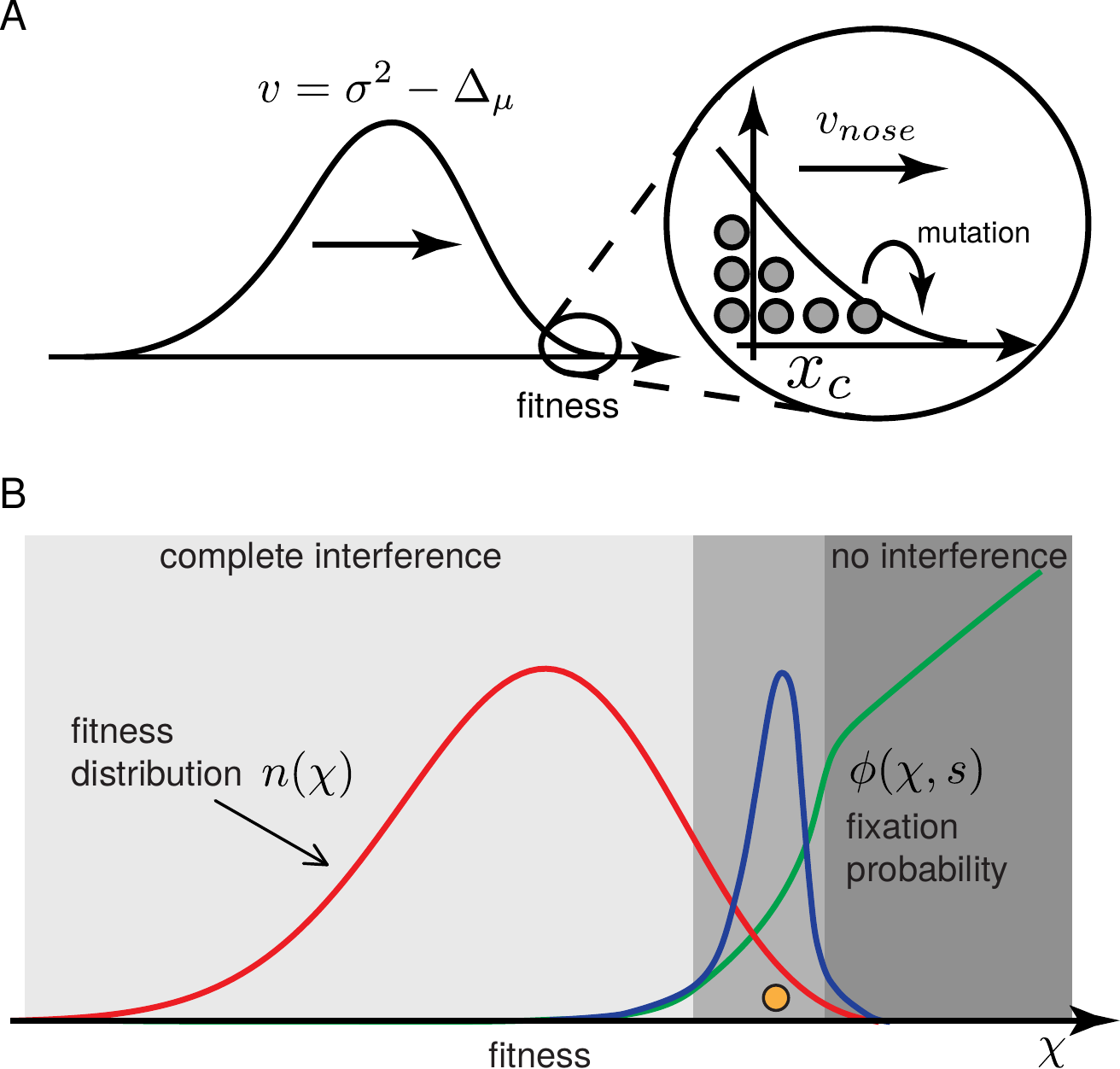}
    \caption{{\bf Traveling wave models} A: The mean fitness in the population increases with rate $\sigma^2 - \Delta_\mu$, where $\sigma^2$ is the variance in fitness and $\Delta_\mu =- u \langle s \rangle$ is the mutation load be generation. The fitness variance $\sigma^2$ is determined by matching the speed of the mean fitness to the speed at which the ``nose" of the fitness distribution advances. The latter is determined by the stochastic dynamics of high fitness individuals. B: The fixation probability (green line) is very small in the bulk of the fitness distribution. In contrast, fixation of a mutation that increases fitness of an individual beyond the nose of the distribution is almost as likely as if there was no competition. The product of the population fitness distribution and the fixation probability defines a narrow zone around $x_c$ where the ancestors of future populations originate from.}
    \label{fig:traveling_wave}
\end{figure}

The earliest models of diverse populations under selection were developed to describe plant and animal breeding. 
In sexual populations with large genomes, quantitative traits typically depend on many loci resulting in an approximately Gaussian distribution of the trait in the population, see Fig.~\ref{fig:traveling_wave}A. 
This model is known as the infinitesimal model with many effectively unlinked loci.
The response to selection will then be proportional to the trait variance and the heritability of the trait the breeder is selecting on \citep{falconer_introduction_2003}.

Outside of plant or animal breeding programs, natural selection operates on fitness. Individuals at the high fitness end of the fitness distribution increase in frequency, while less fit lineages die out.
The dynamics of the bulk of the fitness distribution can often be described by deterministic growth and shrinkage of lineages and the overall rate of adaptation is given by the variance in fitness on which selection can act.
This deterministic dynamics describes the short term response to selection, but it cannot explain how fitness variation is maintained over longer times. 
De-novo mutations that contribute to future fitness variation arise in the high fitness tail of the distribution where only very few individuals reside and stochastic dynamics are important.
Fluctuations and stochastic events among these high fitness individuals are amplified exponentially and dominate the bulk of the population after a delay. 
Almost all individuals that contribute to future populations originate from a narrow zone in the high fitness tail (see Fig.~\ref{fig:traveling_wave}).
Hence a probabilistic description of the high fitness 'nose' is essential.

Tsimring et al.~\cite{Tsimring:1996p19688} studied this problem in a simple model of an adapting RNA virus population: Each individual suffers from many small effect mutations, resulting in a diffusive motion in fitness.
The deterministic diffusion/selection equation describes the dynamics of the bulk of the distribution. 
The stochastic instability in the nose is handled by a cut-off beyond which exponential amplification does not operate.
This ad-hoc modification of the model is justified by the idea that ``fractional'' individuals do not reproduce and no exponential amplification can take place in this part of the distribution.
This modification results in a consistent solution that predicts that fitness variance and the speed at which the population adapts depends only logarithmically on the populations size:
the fittest individuals grow exponentially and hence need only a time $\sim \log N $ to take over the population. Diversity accumulated over this time determines fitness variation.

Tsimring et al \citep{Tsimring:1996p19688} assumed frequent small mutations such that the fitness distribution always remains smooth and no large monomorphic clones exist. 
At steady state, new mutations generate diversity at the same rate as selection removes diversity through amplification of the fittest individuals \citep{neher_genetic_2013}. 
This balance, results in a steady state variance of the fitness distribution
\begin{equation}
    \sigma^2 \sim D^{\frac{2}{3}} (\log ND^{\frac{1}{3}})^{\frac{1}{3}},
\end{equation}
where $D=u\langle s^2\rangle$, $u$ is the total mutation rate, and $\langle s^2\rangle$ is the second moment of the distribution of fitness effects of novel mutations. 
The rate of adaptation, i.e.~the rate at which fitness increases, is then $v=\sigma^2 + u\langle s\rangle$ where $u\langle s\rangle$ is the net effect of mutations.
The fitness increase will typically be offset by a co-evolving population or the environment.

If mutation rates are small, yet populations large enough that many variants compete, other models are a more natural starting points \citep{rouzine_solitary_2003,Desai:2007p954,good_distribution_2012}. 
A commonly used model assumes that fitness can increase in discrete steps of size $s$ (stair-case model) and the rate of such beneficial mutations is $u$. 
In this case, the speed of adaptation increases with $N,u$ and $s$ qualitatively as \citep{Desai:2007p954}
\begin{equation}
  v \sim
        2s^2 \frac{\log Ns}{(\log u/s)^2} \ .
\end{equation}

In both models the speed of adaptation depends logarithmically on the population size and the nose of the wave is $\sim\sqrt{\log N}$ standard deviations above the mean. 
Newly arising mutations have a negligible chance of spreading through the population unless they emerge in the very tip of the fitness wave, see Fig.~\ref{fig:traveling_wave}B.
These features are in stark contrast with traditional models of sequential adaptation, where all individuals equally likely contribute to the future and the rate of adaptation is linear in the mutation rate and population size.

\subsection*{Recombination and adaptation}
The above models were restricted to asexual populations, but traveling wave models can be generalized to sexual populations.
Many eukaryotic organisms recombine by crossing-over, that is combining contiguous stretches of parental chromosomes while maintaining the order of genes and the gene content.
As a result, the rate at which loci are separated via recombination depends on their distance on the chromosome. 
On top of crossing over, different chromosomes are reshuffled by the random partitioning during meiosis.
This sort of reshuffling is also common in segmented viruses like the influenza virus (known as reassortment).
Bacteria exchange genetic material via the exchange of fragments (a few thousand bases long) that either replace a homologous copy or add to the gene repertoire of the bacterium \citep{thomas_mechanisms_2005}. 
All these different modes of recombination can speed up adaptation dramatically. 

\emph{Recombination by reshuffling and reassortment} is easiest to analyze mathematically. 
In this case, the rate of recombination is controlled by the frequency at which individuals reproduce sexually vs asexually.
At very high rates of recombination, the dynamics at different loci in the genome are completely decoupled and adaptation $Nus^2$ is linear in population size and beneficial mutation rate $u$.  
Once the recombination rate drops below typical fitness differentials in the population, clonal subpopulations begin to form and competition between these clones limits adaptation \citep{neher_competition_2009,rouzine_evolution_2005}.
The rate of adaptation increases quadratically with the recombination rate $r$ \citep{neher_rate_2010}:
\begin{equation}
v\approx 
\begin{cases}
\frac{2r^2\log(Nu)}{(\log r/s)^2} & r\ll \sqrt{Nus^2}\\
Nus^2 & r> \sqrt{Nus^2}
\end{cases}
\end{equation}
Whenever the recombination rates are smaller than the typical fitness differences between individuals, the rate of adaptation is insensitive to the rate or effect size of beneficial mutations ($u$ and $s$, respectively) but is instead limited by competition between variants.

\emph{Recombination by crossing-over} implies that distant loci are essentially unlinked, while nearby mutations are inherited together for many generations. 
Recombination is less effective in speeding up adaptation in this case. 
Weissman and Barton \citep{weissman_limits_2012} showed via a scaling argument that the rate of adaptation is linear in the expected number of cross-overs per chromosome when adaptation is dominated by large effect mutations. 
In the opposite limit of many infinitesimal small mutations, the speed of adaptation increases only with the square root of the recombination rate \citep{neher_coalescence_2013}.

\emph{Recombination by horizontal transfer} is different from other modes of recombination that it affects only a small fraction of the genome, similar to mutations.
However, while a {\it de novo} mutation is a random sample from a distribution of mostly deleterious mutations, horizontal transfer will preferentially introduce variation from a successful individual. 
Hence even the exchange of individual loci can dramatically speed up adaptation \citep{cohen_recombination_2005,neher_rate_2010}.


\section*{Genetic diversity in rapidly adapting populations}
Most of the early studies of traveling wave models calculated the speed of adaptation, i.e., the rate at which populations accumulate beneficial variants.
Adaptation, however, is difficult to measure and even when measured has limited power to differentiate models. 
Much more accessible and more informative are patterns of genetic diversity such as the site frequency spectrum (SFS). 
The SFS is the density of mutations found at different frequency in the population. 
Since different models make qualitatively different predictions for the SFS, the SFS can potentially be used to differentiate between models. 

Neutral diversity at a locus is determined by the genealogical tree: the branch lengths are proportional to the number of mutations and the branching pattern determines how common a mutation has become since its origination. 
The simplicity of the Kingman coalescent allows for analytic expressions for many neutral diversity statistics, including the SFS. 

Recently, similar progress has been made for traveling wave models of adapting populations.
Brunet et al.~\citep{Brunet:2006p47336} showed that a class of models, in which a population moves as a front in fitness space gives rise to a coalescent process known as the Bolthausen-Sznitman coalescent (BSC).
The BSC has intriguing connections to the hierarchical structure of ground states in spin glass models, see Ref.~\cite{berestycki2009recent} for a comprehensive review. 
In population models, the BSC emerges naturally if the number of offspring individuals that contribute to the next generation is drawn from a distribution that has a power-law tail $\sim n^{-2}$ (see Ref \citep{schweinsberg_coalescent_2003}).
Such a long tailed distribution has a diverging variance, immediately suggesting that no effective population size can be defined and that the Kimura diffusion equation is inapplicable.

This same process was later shown to describe populations in traveling wave models of rapid adaptation \citep{desai_genetic_2012,neher_genealogies_2013}.
A large class of models seems to converge to the same coalescent model after coarse-graining in time \citep{neher_genealogies_2013}.

The trees generated by the BSC differ qualitatively from those generated by Kingman coalescent trees (Fig.~\ref{fig:coalescent_trees}). 
While Kingman trees tend to branch symmetrically (given $k$ lineages, the distribution of leaves is uniform on the $k-1$ dimensional simplex), BSC trees are often very skewed. 
In Kingman trees, the merger rate is proportional to square of the number of lineages and hence most mergers happen in the recent past.  
In the BSC, coalescence happens deeper in time and several lineages merge at once in multiple mergers.
In traveling wave models of adapting populations, such multiple mergers correspond to exceptionally fit individuals \citep{Brunet:2006p47336,neher_predicting_2014}.

The BSC is an exchangeable coalescent process, that is all individuals in one generation are equivalent and equally likely to contribute to the next generation -- in other words the BSC is a neutral process.
How come genetic diversity in traveling waves of adaptation can be described by such a neutral process?
This effective neutrality only emerges after suitable coarse-graining in time. 
Exceptionally fit clones expand exponentially at slightly different rates, but these rate differences are amplified into a long-tailed distribution of offspring after a characteristic time scale \citep{Brunet:2006p47336}. 
This offspring distribution has a power-law tail $\sim n^{-2}$ and occasionally one lineage takes over a macroscopic fraction of the population.
These rare events correspond to characteristic multiple mergers in the tree. 

Over this same time scale, the fitnesses of lineages competing against each other in the high fitness nose decorrelate and the dynamics become essentially neutral but with anomalous fluctuations.
In essence, heritable variation in fitness (resulting in small systematic differences in offspring number) is exponentially amplified into heavy-tailed iid fluctuations on intermediate time scales. 

\begin{figure}
    \centering
    \includegraphics[width=0.8\textwidth]{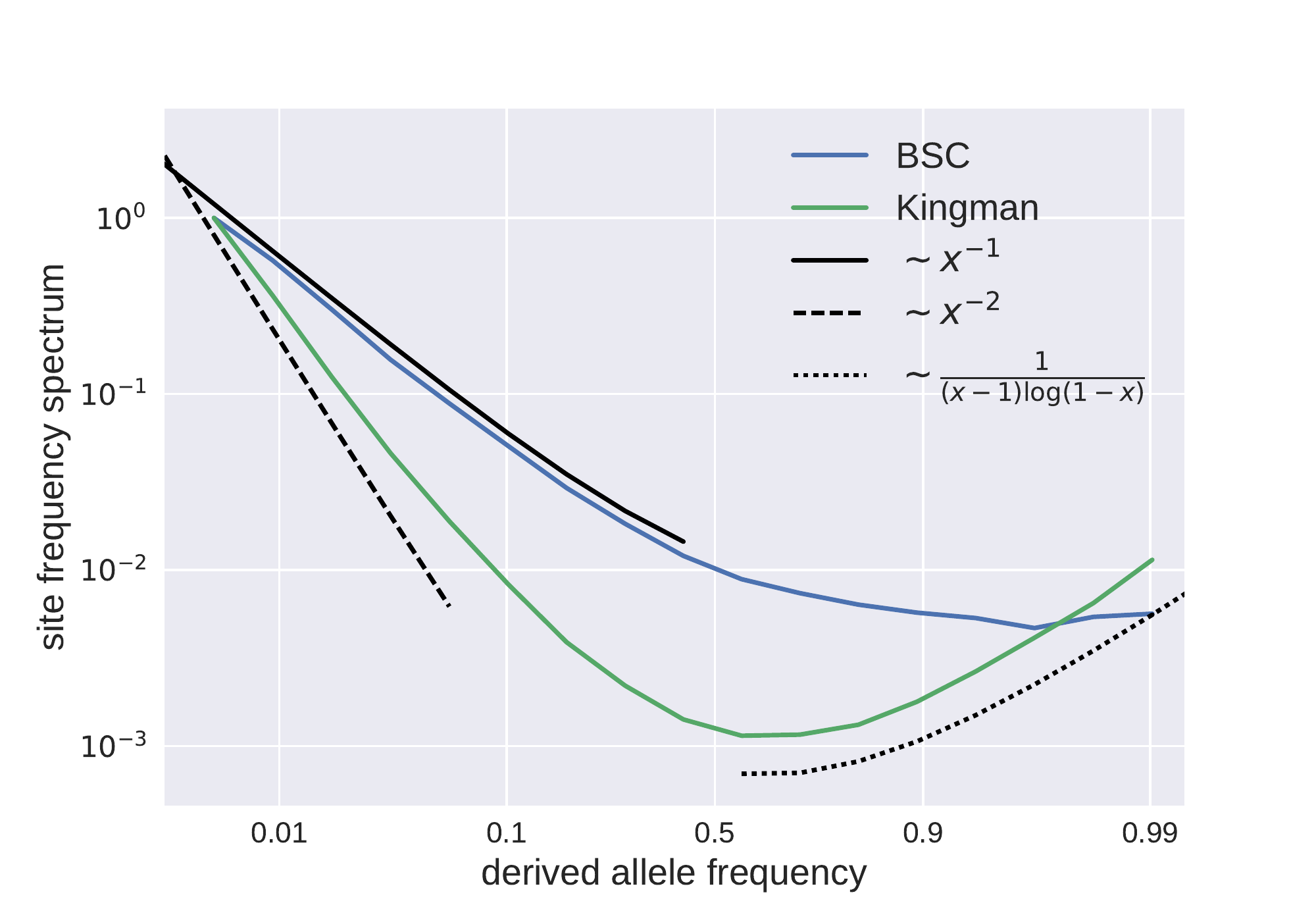}
    \caption{Site frequency spectra differ qualitatively in neutral and rapidly adapting populations. While the SFS in the Kingman coalescent is monotonic and decays as $1/x$, the SFS of the BSC is non-monotonic and diverges again as $(x-1)\log(1-x)$ as the frequency $x$ approaches 1. }
    \label{fig:bsc_kingman_sfs}
\end{figure}

The BSC and the Kingman coalescents make qualitatively different predictions for the SFS (see Fig.~\ref{fig:bsc_kingman_sfs}).
Comparing predictions for distributions like the SFS to data is much more powerful than rejecting null hypothesis based on scalar summary statistics such as Tajima's D. 
Where as the SFS in the Kingman coalescent is monotonically decreasing, the SFS of rapidly adapting populations is non-monotonic with specific asymptotic behavior both for rare and common variants \citep{neher_genealogies_2013,kosheleva_dynamics_2013,hallatschek_selection-like_2017}.
Importantly, the SFS in the Kingman coalescent stays monotonic even if the population size changes through time \citep{sargsyan_coalescent_2008}. 

Qualitative features of the SFS predicted by the BSC have been observed in systems we expect to be under strong positive selection, for example HIV-1 populations of B-cell clones during somatic hypermutation \citep{zanini_population_2016,horns2017signatures,Nourmohammad2018}. Interestingly, the B-cell repertoires of healthy individuals also show signatures of selection in their SFS~\cite{Nourmohammad2018}, showing that SFS can detect the result of many weaker rounds of selection.

The correspondence between the BSC and models of adaptation requires a large number of mutations that contribute similarly to the organism's  fitness. 
More typically, a moderate number of mutations contribute most of the fitness variation, which limits the quantitative agreement between models and data. 
Nonetheless, the BSC as a universal limiting case of rapidly adapting polymorphic populations is a very useful abstraction.

\subsection*{Open problems: genetic diversity in rapidly adapting sexual populations}
In asexual populations, the entire genome is described by one genealogical tree.
Recombination decouples the histories of different loci resulting in a complicated inter-dependent ``forest''.
As one traces the history of a sample of genomes back in time, parts are inherited together for a while until recombination separates linked loci and their histories are described by different trees. 
The ensemble of trees and their connections is known as the ancestral recombination graph (ARG). 
Inferring ARGs from sequence data is an extremely difficult computational problem and most approaches cannot handle more than a few dozen genomes \citep{rasmussen_genome-wide_2014}.

In the limit of very rapid recombination (compared to the time scale of selection), different loci in the genome completely decouple and selection operates on individual alleles rather than individuals or genotypes. 
This reduces the problem to many single locus systems, where residual linkage can be taken into account perturbatively \citep{Neher:2011p45096}.

At lower recombination rates selection and recombination interact in complicated ways and our understanding of genetic diversity in this regime is poor. 
When recombination operates via crossovers, an effective description in terms of effectively asexual blocks captures much of the phenomenology well \citep{neher_coalescence_2013,good_genetic_2014}.
The length of these blocks has to be determined via self-consistent solutions to equations linking recombination, adaptation, and coalescence, but clearly linkage on different distance scales cannot all be lumped into an effective block description. 
Similarly, occasional chromosome reshuffling/reassortment generates BSC-like patterns of diversity \citep{neher_genetic_2011}, but even models that make drastic approximations are not understood.

Even less is known when the epistasis, that is non-additive effects of different loci, interact with recombination. 
Recombination then has the tendency to break up co-adapted combinations, which counteracts the benefits of recombination \citep{neher_competition_2009}, as has for example been observed in the influenza virus \citep{villa_fitness_2017}.

\section*{Repeatability and predictability}
In addition to comparing diversity to model predictions, reproducibility and predictability of evolutionary dynamics can be used to estimate parameters, test models, and interrogate the redundancies at different levels of the genotype-phenotype map. 
At the level of phenotypes, the response to selection is extremely reproducible. Selection on quantitative traits in animals and plants, for example, invariably leads to a strong and consistent response. Anti-microbial resistance evolves rapidly.   
But reproducibility of the accompanying changes in the genotype vary.
Strong very specific selection, for example for resistance to drugs with a well defined target, can lead to parallel mutations in the same nucleotide. This happens frequently in viruses like HIV. 
In bacteria, however, it is more common that drug resistance selection is reproducible at the level of the gene or the biochemical pathway. 
Toprak et al.\citep{toprak_evolutionary_2012}, for example, observed repeated mutations in the same genes of E.~coli in response to antibiotic selection.

One of the biggest systematic studies of repeatability and consistency of molecular evolution was undertaken by Olivier Tenaillon and colleagues \citep{Tenaillon:2012p47907} who selected many initially identical {\it E.~coli} populations to reproduce at elevated temperatures.
This study revealed a striking parallelism in the genes and pathways that accumulated mutations, while repeatability at the individual nucleotide level remained limited. 
Only in cases with very specific selection pressures, like selection for drug resistance, or in organisms with very high mutation rates, are repeated mutations the norm. 

While the repeatability of pathways hit by mutations has been observed in other instances of microbial adaptation, our ability to predict {\it a priori} the genomic responses to particular environmental challenges is essentially non-existing.
Instead of predicting where and how adaptation can occur, one could settle for the more modest aim of predicting which of the existing genotypes will prevail and take over. 
In this case the challenge moves from predicting molecular changes to predicting the changing population composition -- ideally from a one-time observation.
Several such approaches have been developed to predict influenza virus variants that likely dominate future influenza seasons \citep{morris_predictive_2017}.

\L{}uksza and L\"assig \citep{luksza_predictive_2014} developed a model for virus fitness based on scores for mutations that have been historically associated with antigenic novelty and for mutations that are expected to be detrimental, for example because they destabilize protein structure.
Neher et al \citep{neher_predicting_2014} used an alternative approach that infers fitness of different parts of the tree given the tree topology and branch lengths.
This inference is based on a traveling wave fitness model, in which lineages diversify and adapt by many small effect mutations. 
The distinguishing feature of the latter approach is that it only requires a single time point without any influenza specific input or historical data.
The link between tree shape and fitness provided by the traveling wave model allows to infer growth rates without dynamical information. 
This approach therefore can be applied to systems were historical data are not available and little {\it a priori} information of fitness correlates exist \citep{horns2017signatures}.
The approach by \L{}uksza and L\"assig, on the other hand, is easier to complement with additional biophysical or antigenic phenotypes. Both approaches now contribute to the biannual vaccine strain selection. 

\section*{Complex and variable environments}

So far we have assumed that populations adapt towards a well defined and fixed goal, for example rapid growth in evolution experiments or immune evasion in host-pathogen systems. 
More generally, however, the biotic and abiotic environment will change on many time scales and the extent to which lessons learned from studying systems in stable environment are applicable to natural situations is unknown. 
To address this question researchers studied the evolution of bacteria in the mammalian gut. 
The rate of molecular evolution was found to depend strongly on the bacterial strain: experiments with a pathogenic hospital strain showed that adaptation to the mouse gut proceeds $\sim 5$ times slower than in experiments with the laboratory strain K-12 \cite{Lescat2017}. 
This difference likely stems from extensive lab adaptation of K-12 since its isolation in the 1920ies, which rendered K-12 unfit in the gut. 
Starting from a low initial fitness many beneficial mutations are available. Similar patterns have been observed in the lab. 

In analogy to {\it in vitro} experiments, evolution in the mouse gut exhibits convergence among replicate lineages, mainly at the phenotypic, pathway, and gene level \cite{Barroso-Batista2014a, Barroso-Batista2015a, Lescat2017}.
Furthermore, strain with drastically elevated mutation rates, aka mutators, frequently emerged.
While strains with mutator phenotypes have an initial advantage and adapt faster, they accumulate many secondary mutations that are deleterious after the environmental changes and eventually loose out upon transmission and colonisation of new individuals \cite{Giraud2001}. 

Theoretical work that focused on constant environments found that the dynamics and interplay between beneficial and deleterious mutations is key for the survival of mutators \cite{DesaiFisher2011}. Mutators do accumulate many deleterious mutations, but can nevertheless persist at the population level simply because they are continuously produced {\it de novo} and are only gradually eliminated from the population.  Additionally, even though rare, a beneficial mutation generated by the mutators can increase mutator frequency and deleterious mutations can fix in the population through hitchhiking \citep{barrick_genome_2009,doselmann_rapid_2017}. 

Comparison of experiments {\it in vitro} and in the mouse gut suggest that drastic environmental changes often result in early mutations in global regulators. Mutations in global regulators are ways of encoding a large phenotypic shift and are more likely to happen when adapting to new environments.
Such mutations are very common early mutations in laboratory experiments and did occur in experiments colonizing mice guts with the lab adapted K12 strains\cite{Barroso-Batista2015a}, but were not observed in the gut with a pathogenic hospital strain \cite{Lescat2017}. 

Furthermore, interactions of the microbial population with the host are important. 
Molecular evolution of {\it E.~coli} was found to be much slower in immune compromised mice than in healthy wildtype (WT) mice, even though the mutation rates of the bacteria are the same in both environments \cite{Barroso-Batista2015a}. The interaction of the microbiota with the immune system is a fascinating example of co-evolution in complex environments, with practical health implications. 

Theoretical investigations of evolution in fluctuating environments have shown that environmental fluctuations can amplify demographic fluctuations, which results in more rapid coalescence, lower genetic diversity, and a higher probability that deleterious mutations fix compared to a constant environment \cite{mustonen2009fitness,Cvijovic2015,Melbinger2015}.
Fluctuations in the environment translate into large fluctuations in the sizes of the different clones that make up the populations. For example, the long tailed clone size distributions observed in adaptive immune repertoires, measured as the abundances of the B or T-cell receptors in a given individual \cite{Mora2016e, Mora2010}, can be explained in terms of strong fluctuations in the pathogenic and self protein environments the repertoire experiences, which translates  into strong fitness fluctuations felt by each receptor clone \cite{Desponds2016}. 

Organisms can develop strategies to mitigate the effects of fluctuating environments \cite{Kussell2005, Rivoire2016}. These strategies have been studied by considering a population of size $N_t(\sigma)$ at time $t$, where individuals are described by the their phenotype or genotype $\sigma$. The environment changes in time between different phenotypic states denoted by $x_t$. In each generation individuals can switch their phenotype or keep their parents' but they cannot go extinct. Different phenotypes are better suited for different environments, with the best phenotype guaranteeing more offspring and larger fitness. 
Different phenotype switching strategies result in different long term population growth rates, $\Psi[x_{0:t}]= \ln \frac{\sum_{\sigma} N_t(\sigma) }{\sum_{\sigma} N_0(\sigma) }$, and the optimal strategy depends on the environmental fluctuation statistics \cite{Kussell2005,Skanata2016}. 
In extended approaches, differentiating between phenotype and genotype leads to considering different modes of inheritance and sensing \cite{Rivoire2011}. In all cases, the optimal strategies are strongly dependent on the timescales of environmental fluctuations. For example, the diversity of vertebrate and microbial forms of immunity can be rationalized using this theoretical framework in terms of the differences between the timescales of the lifetime of the host compared to the pathogen: vertebrates typically have much longer timescales than their pathogens, while microbe lifetimes are similar to those of their invaders \cite{Mayer2015}. The different immune strategies (such as CRISPR-like vs bet-hedging or adaptive vs innate immunity) then would result from differences in the frequencies of the invading  pathogens.

Lastly, it is also worth noting that these frameworks based on considering long term fitness make direct links with stochastic thermodynamics and information flow \cite{Mustonen2010b, Kobayashi2015a,Sughiyama2015,Kobayashi2015, Rivoire2016}. The long term fitness can be decomposed into ratios of probability distributions that correspond to knowing the environment and the uncertainly coming from the environment. 
While these terms are very hard to evaluate in detail, since they account for path integrals over random trajectories, they correspond to well known fluctuation relations in stochastic thermodynamics, such as the Jarzynski equality \cite{Jarzynski1996, Crooks1998, Crooks1999}. 
Despite certain attempts to link with experiments \cite{Lambert2015, mustonen2006}, this body of work remains mainly theoretical, accentuating the rare event nature of evolution and connecting evolution to ideas in sensing and information flow at the formal level.

\begin{figure}[tb]
    \centering
    \includegraphics[width=\columnwidth]{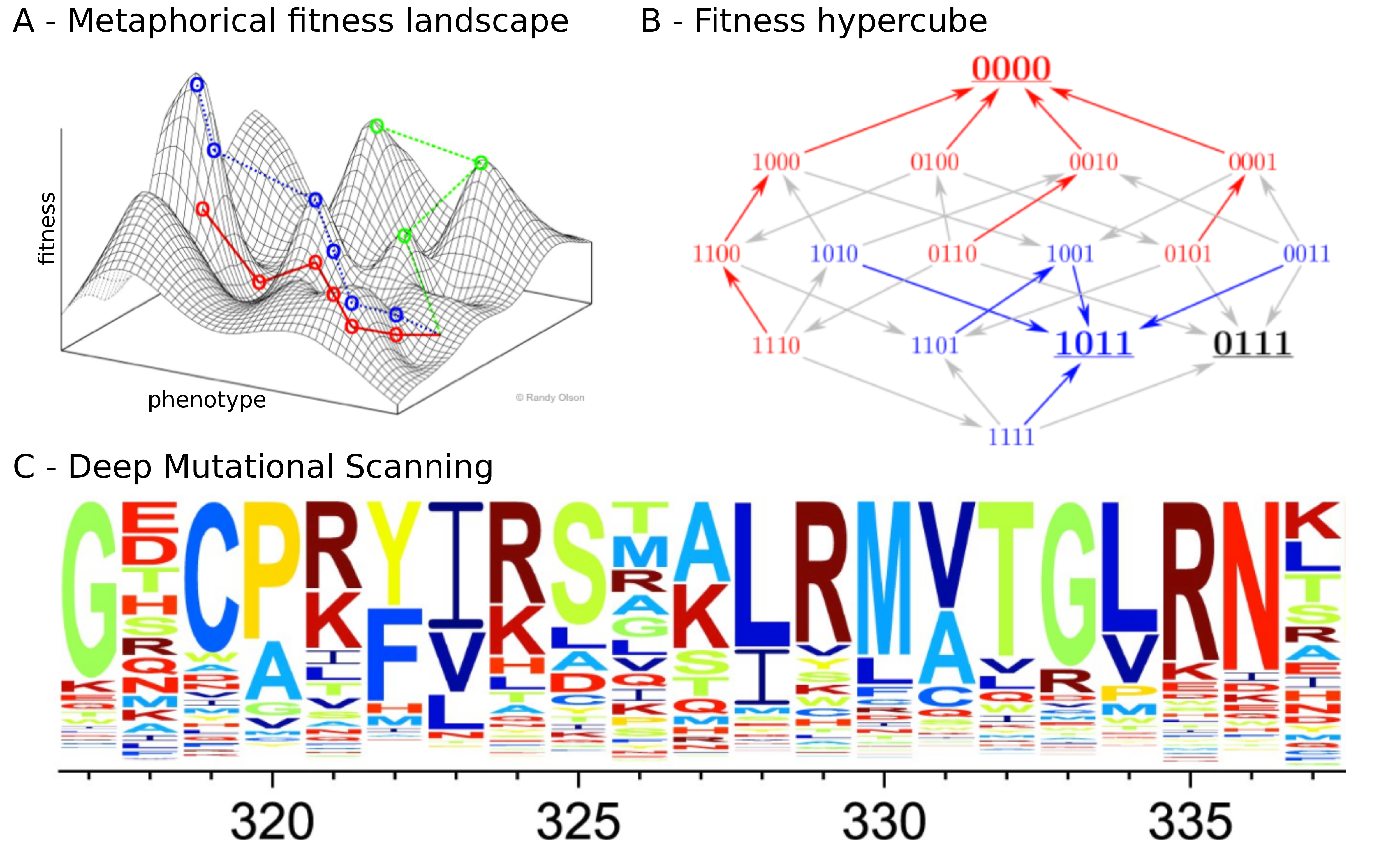}
    \caption{{\bf Fitness Landscapes} A) Fitness landscapes are often drawn in two dimensions where axes represent genotype or phenotype in some abstract away. Mutational paths, illustrated by colored lines, tend to maximize fitness. Depictions of this nature are more metaphorical than faithful descriptions of reality (modified from Randy Olsen, wikipedia). B) Fitness graph hypercubes offer a more informative description of the genotype-phenotype map. A four dimensional hypercube of a fitness landscape in {\it Aspergillus niger} (reproduced from Ref.~\cite{franke_evolutionary_2011}). C) Deep mutational scanning (DMS) experiments quantifies amino acid preferences at each site. Rather than a complete landscape of all combinations of mutations, DMS explores the landscape around a wildtype sequence (this example of mutational tolerance of the influenza virus hemagglutinin protein is reproduced from \citep{thyagarajan_inherent_2014}. }
    \label{fig:fitness_landscapes}
\end{figure}

\section*{Empirical fitness landscapes and adaptive paths}
At any given time, the paths along which a population can adapt depend on the environment and the organism's biology which is a product of billions of years of evolution. 
The effect of individual mutations, as well as which evolutionary paths are accessible, are therefore contingent on this history. 
These constraints on evolution are often conceptualized as ``fitness landscapes''.
Historically, fitness landscapes were depicted in two dimensions as shown in Fig.~\ref{fig:fitness_landscapes}A.
These illustrations are a metaphor, but do not capture the structure of the genotype-phenotype map.
Explicit genotype-phenotype maps are better represented as a fitness function on the hypercube of genotypes where edges link sequences that differ by a single mutation at one position. 
One such hypercube for four specific mutations in the fungus {\it Aspergillus niger} \cite{visser_test_1997,franke_evolutionary_2011} is illustrated in Fig.~\ref{fig:fitness_landscapes}B.
Each pair of sequences that differ by one point mutation are connected by an arrow that points to the genotype of higher fitness. The peaks of the fitness landscape can be identified as sequences with only incoming arrows. 

Weinreich et al.~\citep{Weinreich2006} determined a similar fitness landscape by reconstructing all possible intermediate mutants between inhibitor sensitive and resistant variants of the TEM-1 beta-lactamase.
In contrast to the {\it A.~niger} landscape, these five mutations are in close proximity in the genome and connect known functional end-points.

Five mutations in this enzyme are enough to increase  resistance by $\sim10^6$ fold compared to the wild-type (WT) TEM-1. 
The authors constructed all $5!=120$ mutational paths between the five point mutant and the WT.
Most of the 120 paths passed through strongly deleterious genotypes: only 18 were found to avoid  them and showed up as feasible evolutionary trajectories. Still, half of the weight of all possible evolutionary trajectories followed just two paths. The preference for these paths arises because the order of mutations is important for their effect: mutations that are beneficial in one mutational background become deleterious in another background. . 
Szendro et al.\citep{szendro_quantitative_2013} review empirical fitness landscapes of this type.

In contrast to the above approaches that investigated all possible combinations of a small number of mutations, deep mutational scanning (DMS) experiments \cite{Fowler2010} generate close to all possible single mutants of a wildtype protein and measure their relative performance with respect to a specific function (e.g.~binding of a ligand, enzymatic activity, stability). 
An example of the output of a DMS experiment that represents amino acid preferences at specific positions  is shown in Fig.~\ref{fig:fitness_landscapes}C.

\section*{Inferring fitness and structure from sequence data}
\begin{figure}
    \centering
    \includegraphics{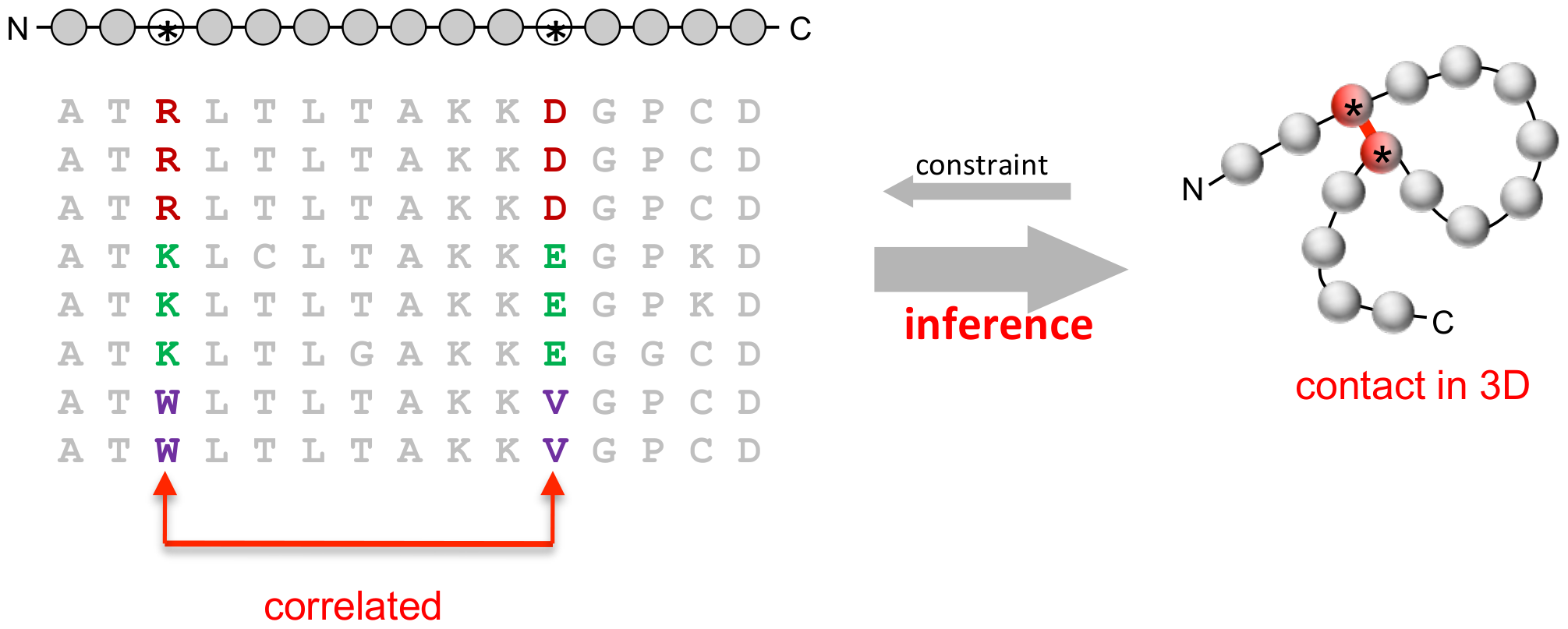}
    \caption{Interacting residues in a protein result in correlated amino-acid substitutions and hence correlated alignment columns. Such correlations in large alignments have been used to infer protein structure and interaction surfaces. (Figure from Ref. \citep{marks_protein_2011}).}
    \label{fig:prot_seqs}
\end{figure}

Since it is easier to destroy protein function or a working pathway than to improve function, the majority of mutations in functional regions of a genome are expected to be deleterious. 
Mutations in these regions are pruned by purifying selection and even distantly related proteins tend to have similar sequence in such regions.
From protein families -- statistically meaningful ensembles of proteins coming from different organisms that can be aligned to each other since they share an evolutionary history -- we can learn how protein structure and function constrain evolution. By analyzing alignments of these families, we see that mutations at certain sites are more common than at others (Fig.~\ref{fig:prot_seqs}). Additionally mutations at certain sites are correlated: mutating one amino acid at a certain position may always be accompanied by a mutation at another position \cite{Neher1994}. In certain protein families correlations between amino acids in their sequences prove extremely important for their structure and function: a substantial fraction of artificially engineered  proteins that constrain the same pairwise correlations as naturally occurring WW domains were shown to fold and have a similar function to WW domains found in nature \cite{Socolich2005, Russ2005}.

These promising results started a whole field of understanding to what degree contacts between amino acids and protein structure can by inferred from pairwise correlations, $C^{ab}_{ij}=f^{ab}_{ij}-f^{a}_if^{b}_j$, between the occurrence of amino acid $a$ at position $i$ and amino acid $b$ at position $j$\cite{BialekRanganathan2007, Weigt2009,Figliuzzi2016}. 
These point $f^{a}_i$ and pairwise $f^{ab}_{ij}$ frequencies can be estimated from data, or learned from a model that describes the probability of the whole protein sequence $\vec{\sigma}$, $ P({\vec \sigma})$. 
Overall, these models and other analysis~\cite{Bloom2005, DePristo2005, Wylie2011, Jacquier2013} of ordered proteins show that protein stability mediated by amino acid interactions is an important determinant of fitness.

A model that maximizes the entropy, $S=-\sum_{{\vec \sigma}} P({\vec \sigma})\log_2 P({\vec \sigma})$, of the possible protein sequences, $\sigma$, while constraining the experimentally measured correlation matrix, $C^{ab}_{ij}$, and imposing normalization of the probability to see a given protein sequence, $\sum_{{\vec \sigma}} P({\vec \sigma})  =1$, leads to a prediction for the form of the probability to observe a given protein sequence, $P({\vec \sigma})\sim \exp[\sum_{i, j, a, b} {J}_{ij} (\sigma^a_i, \sigma^b_j)]$. $\sigma^a_i$ denotes amino acid $a$ at position $i$ and ${J}_{ij} (\sigma^a_i, \sigma^b_j)$ is the $20 \times 20$ matrix encoding the interactions between all possible amino acids at position $i$ and $j$. 

This framework infers pairwise interactions between specific residues, which can then be used to identify interaction surfaces between proteins \citep{Weigt2009} and residue pairs in close proximity in the 3D structure \citep{morcos_direct-coupling_2011}.
Accurate prediction of residue contacts, in turn, allow inference of 3D protein structure from genetic diversity alone \citep{marks_protein_2011}. 

An alternative approach has also been developed based on a delocalized spectral  decomposition of the correlation matrix weighted by positional conservation, $\tilde C^{ab}_{ij}=\sum_k \ket{k} \lambda_k \bra{k}$, into modes $\ket{k}$  termed ``sectors''  ordered by their eigenvalues $\lambda_k$ \cite{Halabi2009, Rivoire2013}. Positions contributing to dominant ``sectors'' are contiguous in the tertiary structure, show common biophysical and biochemical properties and form functional units with independent patterns of sequence divergence. These delocalized and localized frameworks have been unified by showing that they are both limiting cases of a more general decomposition in terms of a Hopfield-Potts model, $P(\vec{\sigma})=\frac{1}{Z}\exp\left[-\sum_{ij} \left(\sum_{\mu=1}^K \lambda_{\mu} \xi_i^{\mu} \xi_j^{\mu} \right) \sigma_i \sigma_j\right]$, where $ \xi_i^{\mu} $ correspond to a set of delocalized modes that can be mapped onto sectors, while their summation reproduces the effective interaction matrix \cite{Cocco2013, Rivoire2013}. For a full description of protein sequence diversity both small and large scales are needed, although often reduced descriptions do very well.

Knowing that correlations are important motivated researchers to map out the mutational landscape and link specific mutations, and their interactions to phenotypic variables, such as the fitness of the {\it E. coli} bacterium carrying a given beta-lactamase TEM-1 mutant \cite{Figliuzzi2016}, or the affinity of antibodies - proteins involved in an immune response - against a specific antigen \cite{Adams2016}. These studies, which are examples of deep mutational scans, find that most mutants are detrimental -- they reduce the fitness or binding affinity compared to the wild-type protein. They also show the importance of amino-acid interactions -- termed epistasis, which is not limited to a few strongly interacting pairs but can take the form of many weakly interacting sites~\cite{Adams2016}. 
The epistatic sites contribute substantially to to antibody binding and 
many of the mutations display positive epistasis~\cite{Adams2017}, potentially enlarging the mutational accessibility of high-affinity antibodies. 

\section*{Evolving novel functions}
Deep mutational scanning experiments probe the neighborhood of a particular wild-type genotype, but typically have little to say about plausible evolutionary paths that are several mutations in length because of epistatic interactions between successive mutations.

Possibly more representative evolutionary trajectories are generated by repeated diversification and selection. 
Similar to DMS, directed evolution experiments generate a diverse library of genotypes and select the best molecules for a given function, such as binding a specific ligand. 
The pool of sequences that emerges as ``winners" can be analyzed to identify successful mutations and serves as a starting point for the next round of diversification. This artificial selection process is very efficient, but it is unclear whether this combination of very high mutation and strong selection results in similar trajectories as evolution in the wild.

Bloom and Arnold allowed the cytochrome P450 enzyme to accumulate mutations that were neutral for one specific substrate \cite{Bloom2007}. When they tested the activity of the mutant enzymes to five different substrates they found that some of these mutants showed up to four-fold higher activity to the new substrates than to the original one. The enzyme evolved what is known as cross-reactivity or promiscuity.
It has been argued that enzyme promiscuity is a critical starting point for evolution of novel activities \citep{copley_shining_2017}.

Affinity maturation in B-cells -- important actors of the immune system that recognize pathogens -- is a natural process that is similar to directed evolution experiments. Affinity maturation starts from a diverse source of immune cells that get selected for their ability to recognize an invading pathogen. The cells that bind more strongly to the pathogenic molecules proliferate faster, undergoing Darwinian evolution: the offspring cells acquire somatic mutations, which then get selected upon in the next round. This way, affinity maturation consists of several rounds of mutations and selection, forming evolutionary lineages just like viruses or microbes, and the affinity of B-cells to pathogenic molecules increases 10-100 fold \cite{Murphy2007}.  At the end, a fraction of the evolved cells are kept in a subcompartment of the immune system called the memory repertoire, which responds rapidly upon re-infection by a similar pathogen, conferring fast protection. Recent experiments have shown that when B-cells are stimulated with a synthetic molecule, a hapten, affinity maturation looks very much like the directed evolution experiment: the B-cell binding affinity for the hapten increases and the diversity of B-cells decreases during the process. However, when B-cells are stimulated with pathogen derived molecules (antigens), binding affinity still increases but diversity remains high \cite{Kuraoka2016}. Interestingly, the cells that are kept in the memory pool at the end of affinity maturation are not necessarily the best binders. So it seems that the real life ``directed evolution" experiment is more complex than the in-lab version. Of course, natural antigens are much more complex than haptens, since they are composed of many receptor binding regions, termed epitopes. Yet the diversity of epitopes is not the only reason for a diverse response. The strength of the response of different selected clones to a specific epitope can vary by orders of magnitude, with even clones that do not strongly bind the antigen surviving affinity maturation. The diversity of B-cells during affinity maturation also increases with time, as initially rare but better binding clones increase in frequency adding to the frequent initial winners. The detailed experiments exclude obvious explanations such as spatial mixing or recruitment of new naive cells, leaving an open question about how a broad response is up-kept and what are its benefits for protection.

\section*{Co-evolution}

Analysis of empirical fitness landscapes showed that evolutionary paths within isolated proteins are strongly constrained. 
Additionally, evolution occurs in the presence of other interaction partners, constraining the fitness landscape of interest by fitness landscapes of interacting organisms and making it dynamic.  

\begin{figure}
    \centering
    \includegraphics[width=0.7\columnwidth]{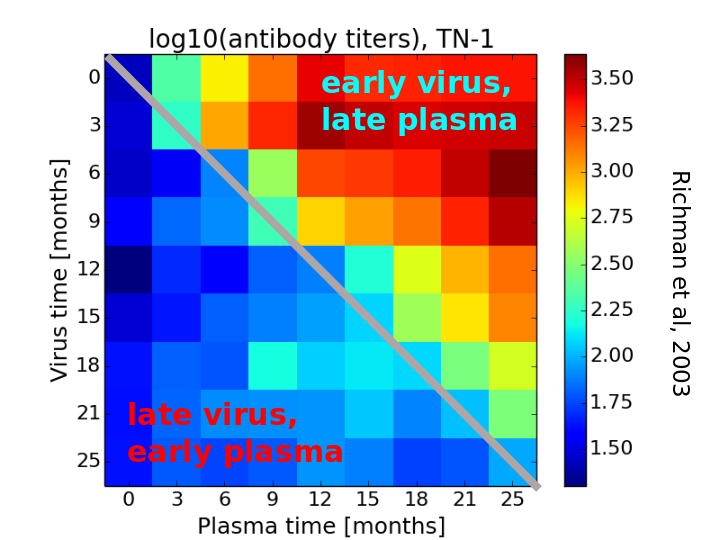}
    \caption{Neutralization titers of longitudinally sampled plasma/HIV combinations (data from \citep{richman_rapid_2003}). Virus is efficiently neutralized by antibodies in plasma sampled 6 month later, but plasma typically does not neutralize virus from the same time point.} 
    \label{fig:patho_coevolution}
\end{figure}

Co-evolution can be readily observed between pathogens and the adaptive immune system. The adaptive immune system consists of a diverse set of B and T-cells that are endowed with specialized receptors able to bind pathogenic molecules, successfully recognize them as foreign to the host organism and trigger an immune response aimed at eliminating them. The diversity of different cells is needed, given the different pathogens a host may encounter. However, as the immune system recognizes pathogens, it exerts pressure on the pathogens that mutate and  escape the immune system. In turn this forces the immune system to evolve and chase the pathogens, resulting in an arms-race like co-evolutionary process. 
Fig.~\ref{fig:patho_coevolution} shows on example of this process from a chronic HIV infection. Antibodies in plasma neutralize virus variants that circulated in the past, but the virus population rapidly evolves new variants that escape neutralization.
Another good example is the evolution of broadly neutralizing anti-bodies against HIV \citep{liao_co-evolution_2013}. Repeated exposure to changing antigens eventually result in antibodies that recognize conserved in-variable features of the virus. 
Similarly, seasonal influenza viruses evade immunity that gradually builds up in the human population \citep{petrova_evolution_2017}.

It is very hard to sample all the possible pathogens that we are exposed to, but thanks to recent high-throughput sequencing technology we can now sample a significant fraction of the immune receptors in a given organism \cite{Weinstein2009, Georgiou2014, Robins2009,Warren2013a,Six2013, Shugay2014,Egorov2015}. These experiments combined with an advanced statistical analysis show that non-related healthy people that do not necessarily share the same lifestyle, share as many unique immune receptors as expected by chance, if we use the structure of the repertoire for the random estimate \cite{Elhanati2014, Pogorelyy2016}. Outliers from these estimates can be linked to functional selection pressures \cite{Pogorelyy2017b}. Even in such diverse ensembles we can thus identify reproducible patterns also at the genotype level and ask how is diversity constrained, given an appropriate statistical analysis.

Ultimately, the convergence of responses matters at the phenotype level: there are probably many different ways of conferring resistance against a given pathogen. Antigenic landscape maps and neutralization assays are examples of phenotypic assays that address this problem~\cite{richman_rapid_2003, Fonville2014}, especially for viral evolution (Fig.~\ref{fig:patho_coevolution}). Different virus strains are exposed to the aliquots from the same blood sample. The ability of the sample to produce protective antibodies is translated into a measure of similarity of viral strains. Lately, high-throughput affinity assays that use specific antigens to ``fish out" responding antibodies provide different phenotypic measurements. Linking different phenotypic levels and linking phenotype to genotype remains challenging, although it is necessary if we want describe co-evolution at the genotypic scale.

Co-evolution is not only important for pairs of organisms, but is also observed between interacting components of cells.  
A number of studies describe how two mutating  molecular partners influence each others evolutionary paths, creating a mutual fitness landscape \cite{Poelwijk2007, McKeown2014, Anderson2015}. Anderson et al.~reconstructed all possible paths in an ancient transition between two hormone ligands and two binding sites, closed evolutionary paths were opened by permissive mutations occurring in the partner. These mutations allowed the first partner to tolerate mutations that would otherwise be deleterious \citep{Anderson2015}. Similarly, certain paths were closed by co-evolution. The evolutionary history of the pair depends strongly on the partners history and shows that co-evolution has strong implications on the space of evolutionary paths. Evolutionary paths, in turn, are trajectories conditional on their surroundings. 

The hormone-binding site study shows an example of the positive effects of co-evolution opening otherwise certain forbidden paths. However, co-evolution can also involve more than two interacting partners that infer strong negative selection pressures upon each other. 

\subsection*{Open problems: constrained viral--immune co-evolution and predictability}
The immune system is an example of a fast evolving system, where the frequencies of different receptor species change with time, responding to changes in the antigenic environments. New ``species" (receptor variants) are constantly trying to ``invade" an the existing population, however their success, or fitness, depends on both the current state of the existing repertoire and the current state of the external environment. The adaptive system functions as a distributed network, and current evidence shows that even intuitive processes such as affinity maturation of B-cells to specific antigens proof to be more complex than we expected. Yet the repeatability of these evolutionary experiments in many organisms gives an opportunity to ask questions that bridge ecology and evolution and quantitatively test theoretical predictions. It is also a good system to push the limits of our ability to predict responses.  Current predictions of future virus strains pick winners from candidate strains that are already present at relatively high frequencies. Formulating the problem in this way, is extremely useful. Yet an ideal understanding of viral and immune co-evolution should be able to {\it a priori} identify new mutations, albeit probabilistically. This is of course an extremely challenging task.  Whenever it fails, it will probably teach us a lot about the limits of predictability, potentially crystallising natural boundaries of these processes.

\section*{Ecology}

Each species evolves in a dynamic environment shaped not only by abiotic factors but also by other species. This naturally generalizes co-evolution to a larger number of interacting species. 
Thanks to the advances in sequencing technologies, we can now sample the diversity of microbial communities in the wild, on fermented food or cheese, and in various bodily orifices \cite{Jovel2016}. We can first ask about the diversity of species living in a given community. The study of different cheese rinds reveals that most of the large bacterial diversity originates from environmental sources, such as the air, as opposed to starter cultures, and that abiotic factors (e.g. salinity, pressure, moisture) shape the community composition \cite{Wolfe2014}. The environment favors certain strains and the growth of these strains changes the environment for the community. Theoretical descriptions of such interactions go back to the 1970s \cite{mac1969species} and are described by generalized consumer--resource, or Lotka-Volterra like models. The differences in the assumptions matter, but in general one considers the change in time of a subpopulation of size $N_i$, due to effective growth $f_i(\vec{N})$, community interactions with other substrains and itself,  $\alpha_{ij}$, and noise, $\xi(t)$: 
\begin{equation}
    \partial_t N_i = N_i( f_i(\vec{N})-\sum_j \alpha_{ij}   N_j) +\xi_i(t) \ .
\end{equation}
Both the effective growth and community interaction terms can additionally depend directly on environmental factors. These approaches allow us to ask questions about timescales, the role of interactions and the environment, and have produced a very large body of work \cite{azaele_statistical_2016, Posfai2017,Haerter2018}. However until recently \cite{tikhonov2015interpreting, Fisher2017} they remained mostly disconnected from experiments. 

Characterizing diversity gives us a static picture of a community. We can also ask how diversity evolves. From studies of the marine plankton bacterium Prochlorococcus we learn that the changes in the environment can be transmitted to the community through changes in the metabolism of its members \cite{Biller2014}. Prochlorococcus lives at different depths in the ocean and its subpopulations have adapted to the levels of light and nutrients available at each immersion depth. The whole population responds to modifications in the external environment, such as light exposure, by a niche constructing chain of adaptation. New ecotypes change their metabolism, chemically modify their environment, which exerts pressure on their co-inhabitants and globally change the diversity of the whole ecosystem\cite{Braakman2017}. Theoretical studies of interacting Lotka-Volterra like models with random interactions show that these systems have a limited capacity for how many different species they can support \cite{mac1969species}  and diversity is predicted to exactly surf the allowed bound \cite{Biroli2017}: a community with initially more species will decrease the number of species until the maximum allowed number survive. This diversity constraint means the system is marginally stable, which is known in spin glass physics to correspond to a highly rugged landscape with many minima. Even small perturbations of the system will lead to large, nonlinear rearrangements of the landscape and a new community composition. These experimental and theoretical results open up a series of questions about the timescales for community rearrangements, their dynamics and whether the communities we observe are stable or transient \citep{kessler2015generalized}. How should we interpret the fact that random interaction models often give reasonable expectations?

The adaptation of Prochlorococcus to different environmental niches is also visible at the level of genetic variation. The genetic diversity is huge: in 1ml of water there are hundreds subpopulations and the sequencing of each new organism adds 160 new genes, which corresponds to $\sim 6-8 \%$ of the known pan genome \cite{Biller2014}. The genetic variation can be linked to niche adaptation. On top of the backbone genome adapted to light, nutrient and temperature variability and shared between organisms in a given niche, each bacterium has many flexible genes, resulting in this very large genomic diversity. Despite large population sizes of $N=10^{13}$, a relatively average mutation rate $\mu=10^{-7}$, but extremely strong environmental selection, no selective sweeps have been reported and the diversity is huge. These observations defy the predictions of traditional population genetics. Why is there so much diversity given the strong selection pressures? Many ideas have been put forward, including co-evolution with predators (phages), recombination, which is extremely common in bacteria in the wild, the potentially complicated role of selective environments, as well as the role of the population structure, which suggest that we should study the whole federation of organisms living in a given ecosystem as selectable units. At this point the list of possibilities means we do yet have an answer.

Cyanobacteria are part of the Prochlorococcus federation. But they also live in hotsprings, where similarly to Prochlorococcus they show huge genetic diversity despite strong selection \cite{Rosen2015}. However, unlike Prochlorococcus, which despite the within-niche diversity does show a niche structure, these cyanobacterial biofilm genomes form a freely recombining gene pool, where allele frequency correlations between individuals decay on very small length scales along the genome. Selection seems to structure the genomes on short genomic scales, while recombination dominates on longer scales. Only an advanced statistical analysis combined with appropriate sequencing experiments was able to reveal this picture of a community without coherent strain that correspond to genome-wide stable ecotypes.

\subsection*{Open problems: linking ecology and evolution}
Linking ecology and evolution is the least developed of the open problems discussed in this review. Similarly to  cyanobacteria and {\it Prochlorococcus}, analysis of {\it E.~Coli} genomes in the wild finds even many interesting signatures of the interplay of the underlying evolutionary forces with the surrounding habitat. These studies leave us with the impression that simply gathering data will not answer the question about the structure of diversity and the relevance of different forces. Only analysis methods driven by a theoretical understanding of the evolutionary processes can help us make sense of the vast datasets and understand the structure of microbial communities. In addition to a lack of understanding of the relevant organizational scales, we know little about the relevant time scales. 
Is the diversity we observe stable or inherently transient? 
Is change driven by properties of the dynamical system, or are systems adiabatically coupled to changing environments?

\section*{Conclusions}
Our understanding of short term evolution in controlled or stereotypic environments has improved dramatically over the past decades. 
At the same time, quantitative high-throughput methods to survey fitness landscapes and eco-evolutionary dynamics of populations expose the complexity of biology and highlight the short-comings of the theoretical models of evolution.

Detailed studies of evolution in microbial populations \citep{Tenaillon:2012p47907,zanini_population_2016,good_dynamics_2017} are first steps towards a more comprehensive understanding of rapid adaptive evolution and we have obtained partial answers to questions like (i) what are the relevant parameters, (ii) how gradual is evolution, (iii) and in what sense and to what degree can evolution be reproducible or predictable \cite{Gould1989, Orgogozo2015, Lassig2017}.
Whether these insights extrapolate to other systems and to different temporal or spatial scale is unclear and doubtful.

In addition to evolutionary dynamics, we have made substantial progress in understanding how proteins or molecular circuits evolve to acquire new functions and what are the constraints on diversification. 
Deep mutational scanning experiments \citep{Fowler2010} allow high-throughput characterization of protein function landscapes, directed evolution experiments have shown how proteins can acquire new functions, and sequencing of immune receptor repertoires has shown how the vertebrate immune systems prepares for future challenges and optimizes its response against recurrent challenges. 
Beyond individual proteins, however, predicting or even characterizing the perturbations remains challenging beyond the crudest of perturbations (e.g.~transposon sequencing knock-out libraries \cite{van_opijnen_transposon_2013}). 

Genetic diversity is easy to quantify these days by high throughput sequencing. 
But interpreting this diversity is difficult since (i) it remains difficult to associate genetic diversity with function and phenotype, (ii) most data are static, and (iii) limited to a minority of the interacting entities. 

Recent metagenomic studies quantifying diversity, composition, and dynamics of microbial ecosystems are just scratching the surface and we have little understanding of the rules governing such communities.
We tried to give examples and show how diversity is observed on many scales and also evolution acts across scales, from the molecular, through the genomic and organismal to the ecological. 
Despite a lot of experimental data and different theoretical approaches that lead to fascinating observations we still lack a unified framework for understanding diversity.
Clearly, as evidenced by microbial communities, the events on one scale influence the observed diversity at another scale. 
Do we need to develop a detailed understanding on all scales to describe a given phenomenon or to can we understand one scale without considering the others? Is there a well defined separation of scales when describing diversity, is everything deeply interconnected or is there are cascade of scales where one scales feeds into the next?

In summary, we have developed a reasonable understanding of dynamics on short time scales in systems where the ``objective'' of the evolving population is clear and static, e.g. the global influenza virus population. 
But our ability to generalize and extrapolate to longer time scales, more complex environments or ecosystems is limited to absent. 

\section*{Acknowledgement}
We are grateful to Benjamin Good and Thierry Mora for critical reading of the manuscript and providing valuable feedback.

\bibliography{solvay,thierry,review}

\end{document}